\documentclass[%
preprint,
amsmath,amssymb,
aps,
]{revtex4-1}

\usepackage{graphicx}
\usepackage{dcolumn}
\usepackage{bm}

\usepackage{textcomp} 
\usepackage{braket} 

\begin{document}


\title{Mutually synchronized macroscopic Josephson oscillations demonstrated by polarization analysis of superconducting terahertz emitters}

\author{M. Tsujimoto$^{1,2}$}%
 \altaffiliation[Corresponding author: ]{tsujimoto@ims.tsukuba.ac.jp}

\author{S. Fujita$^3$}
\author{G. Kuwano$^2$}
\author{K. Maeda$^3$}
\author{A. Elarabi$^3$}
 \altaffiliation[Present address: ]{Okinawa Institute of Science and Technology Graduate University, Onna, Okinawa 904-0495, Japan}
\author{J. Hawecker$^4$}
\author{J. Tignon$^4$}
\author{J. Mangeney$^4$}
\author{S. S. Dhillon$^4$} 
\author{I. Kakeya$^3$}
 \altaffiliation[Corresponding author: ]{kakeya@kuee.kyoto-u.ac.jp}

\affiliation{%
$^1$Faculty of Pure and Applied Sciences, University of Tsukuba, 1-1-1 Tennodai, Tsukuba, Ibaraki 305-8573, Japan
}%

\affiliation{%
$^2$Graduate School of Pure and Applied Sciences, University of Tsukuba, 1-1-1 Tennodai, Tsukuba, Ibaraki 305-8573, Japan
}%

\affiliation{%
$^3$Department of Electronic Science and Engineering, Kyoto University, Nishikyo-ku, Kyoto 615-8510, Japan
}%

\affiliation{%
$^4$Laboratoire de Physique de l’Ecole normale sup\'{e}rieure, ENS, Universit\'{e} PSL, CNRS, Sorbonne Universit\'{e}, Universit\'{e} de Paris, F-75005 Paris, France
}%

\date{\today}

\begin{abstract}
We demonstrate mutual synchronization of Josephson oscillations in multiple stacks of intrinsic Josephson junctions of the cuprate superconductor Bi$_2$Sr$_2$CaCu$_2$O$_{8+\delta }$.  Detailed analysis of the full polarization parameters allows the determination of a phase correlation between the stacks: a simultaneous emission state is described by a linear combination of individual emission states with a phase retardation.  This proves that the stacks are coupled via a Josephson plasma in a superconducting substrate and the coupling matrices can be extracted from polarization analyses.  Our findings suggest a route towards the realization of high-power terahertz sources based on the synchronization of a large number of intrinsic Josephson junctions. 
\end{abstract}

\maketitle


\hrulefill
\vspace{1cm}

The observation of electromagnetic (EM) radiation in the terahertz frequency range emitted from a stack (mesa) of intrinsic Josephson junctions (IJJs)~\cite{Kleiner1992} in Bi$_2$Sr$_2$CaCu$_2$O$_{8+\delta }$ (Bi-2212) highlights the potential of such materials in  monolithically fabricated solid-state terahertz sources~\cite{Ozyuzer2007,Welp2013,Kakeya2016,Kashiwagi2017a}.  The mutual synchronization of IJJ mesas has been demonstrated to produce output powers of up to 0.6~mW~\cite{Benseman2013b}, where the maximum output power exceeds the sum of the individual emission powers.  Although numerical study suggests a possible coupling between the mesas through the Bi-2212 substrate~\cite{Lin2013a}, there is no direct experimental evidence showing the mutual synchronization.  So far, to investigate these phenomena with multi-degree-of-freedom, the frequency and intensity of the emitted EM wave have been considered good measures.  Besides these characteristics, the polarization includes rich information for characterizing the Josephson plasma wave (JPW) polarized along the $c$-axis inside a mesa~\cite{Elarabi2017a,Elarabi2018}.  Since a JPW is converted to a polarized photon according to the mesa geometry~\cite{Koyama2009,Hu2009,Savelev2010a,Klemm2010b}, perfect polarization is the hallmark of coherent excitation of the JPW.

In this study, we demonstrate synchronization of macroscopic Josephson oscillations in two simultaneously biased mesas coupled via a superconducting base crystal by measuring complete Stokes parameters.  Our focus is on the polarization of photons emitted from individual mesas and arrays of mesas.  Most importantly, we observed a drastic change in the axial ratio for the simultaneous emission.  This suggests the possibility of active control of the synchronization in mesa arrays, which is the most promising way to increase the integrated output power.

Mesas 100$\times $400~\textmu m$^2$ in size with two silver strip electrodes were milled from a Bi-2212 crystal by photolithography and argon milling methods. Figure 1(a) shows an optical microscopy image of the Bi-2212 mesa array. We refer to the two mesas as A1 and A2. Profile measurements using atomic force microscopy (Keyence Corp., Model VN-8000) demonstrate that the mesas vary marginally in size. Widths of A1 and A2 were measured as 94~\textmu m and 91~\textmu m, respectively, with thicknesses of 1.4~\textmu m corresponding to 910 IJJs.

The current-voltage characteristics (IVCs) for A1 and A2 show the large hysteresis typical of underdamped Josephson junctions~\cite{SM}.  Simultaneous emission occurs when A1 and A2 are biased in parallel.  Hereinafter, we refer to the parallel connection as A1$\parallel $A2.  The maximum intensity is obtained at 16.8~mA, which is higher than the sum of the bias currents for the maximum emission powers of A1 and A2 individually.  Also, the maximum intensity for A1$\parallel $A2 was almost half of that for A2.  This can be explained by considering the local temperature increase~\cite{Wang2009,Wang2010,Gross2012,Benseman2013c,Minami2014,Tsujimoto2014,Benseman2015}.  Benseman {\it et al.} demonstrated that the self-heating effect limits the power output and in fact may prevent synchronization among multiple mesas~\cite{Benseman2013b}.  A variety of studies on the cavity resonance effect have demonstrated that spontaneous synchronization among stacked IJJs is accompanied by the formation of standing EM waves inside the Bi-2212 mesa~\cite{Ozyuzer2007,Tsujimoto2010a,Kashiwagi2011,Tsujimoto2016,Kashiwagi2018,Zhang2019}.  For a thin rectangular mesa of width $w$ and length $\ell $, the cavity frequency for a transverse magnetic $(mp)$ mode is given by $f_{mp}^{r}=(c_{0} / 2 n)\sqrt{(m/w)^{2}+(p/\ell )^{2}}$, where the two indices $m$ and $p$ correspond to the numbers of electric field nodes in the width and length directions, respectively, and $n = 4.2$ is the experimentally obtained refractive index~\cite{Tsujimoto2016,Kadowaki2010}. Here, the radiation frequencies measured using Fourier transform interferometry ranged continuously from 0.56 to 0.66~THz depending on the bias point~\cite{SM}.  These values are in good agreement with the calculated $f_{1p}^{r}$ with $0<p<6$. If we assume $m \geq 2$, the calculated $f_{mp}^{r}$ values fail to coincide with the observed values for any $p$. Hence, only one EM node is present along the mesa width, and a non-zero $p$ value is expected to produce an elliptical polarization. 

Here, we present the measurements of the Stokes parameters, which allow for the quantitative analysis of polarized photons emitted from individual mesas and from synchronized arrays~\cite{SM}.  Figure 1(b) shows a schematic view of the synchronized array.  Measurement was performed by allowing the polarized radiation to propagate sequentially through two polarizing elements, a quarter-wave plate (QWP), and a linear wire grid polarizer (WGP).  The QWP consists of a stack of parallel metal-plate waveguides~\cite{Nagai2015}.  We used terahertz time-domain spectroscopy~\cite{Madeo2010,Maussang2016} to verify that the QWP works correctly in the emission frequency range around 0.6~THz~\cite{SM}.  Figure 1(c) shows the measurement configuration.  The QWP can be rotated through an angle $\theta $ and is followed by a fixed WGP whose transmission axis is fixed in the width direction ($\theta =0$~deg).

Figures 2(a) and 2(b) show polar plots of the bolometer output for A1 and A1$\parallel $A2, respectively, as a function of $\theta $ at the bias conditions that result in the maximum output powers.  The error bars in the radial direction correspond to fluctuations in the bolometer output signal, mostly owing to background noise.  The four independent Stokes parameters, $S_0$, $S_1$, $S_2$, and $S_3$, which are summarized in Table SI~\cite{SM}, are obtained from this data.  The solid lines shown in Figs.~2(a) and 2(b) represent the calculated results using the four Stokes parameters.  The experimental data are slightly asymmetric with respect to both the major and minor axes.  This is due to the imperfect alignment of the parallel metal-plate waveguides in the QWP.  Nevertheless, the calculation results fit the experimental data within the error bars. 

The E-field vector at the detection plane is given by $\bm{B}(t)=E_{0x}e^{i(\omega t + \delta _x)}\bm{i} + E_{0y}e^{i(\omega t + \delta _y)}\bm{j}$, where the $x$- and $y$-axes are parallel to the mesa width and length, respectively, $t$ represents time, $E_{0x}$ and $E_{0y}$ are the respective amplitudes, and $\delta _x$ and $\delta _y$ are the respective phase constants.  Figures 2(c) and 2(d) are the corresponding polarization ellipses for A1 (red), A2 (blue), and A1$\parallel $A2 (green).  The fourth Stokes parameter $S_3$ determines the helicity of the photons: a positive (negative) $S_3$ indicates left (right)-handed helicity with respect to the direction of propagation.  Note that the $E$-field rotates forward (counter-)clockwise for left (right)-handed helicity from the viewpoint of the detector.  The arrows on the ellipses indicate the direction of the $E$-field rotation.  For both mesas, the major axis of the polarization ellipse is oriented along the $x$ axis, {\it i.e.}, $-\pi /4 < \psi < \pi /4$, which is consistent with excitation of $(1p)$ cavity modes. 

In the pioneering study of EM-wave emission from an IJJ mesa, it is demonstrated that the emission from rectangular mesas is linearly polarized along the mesa width~\cite{Ozyuzer2007}.  The present results suggest, however, that the emitted waves are elliptically polarized with a finite axial ratio at an arbitrary orientation angle.  We stress that these polarization parameters contain information essential for understanding the electromagnetism inside the emitting mesa.  For example, the orientation angle $\psi $ is derived from the phase difference $\delta _{xy} = \delta _{x}-\delta _{y}$, namely, $\delta _{yx}= \pm \pi /2$ gives $\psi = 0$ and $\delta _{yx}=0$ gives $\psi = 30$~deg for  $E_{0x}^2 / E_{0y}^2 \sim 3$. We also found that the actual polarization parameters are, in fact, dependent on the bias condition and $T_b$.  Nevertheless, the observed $E_{0x}$ was greater than $E_{0y}$ in all cases, directly suggesting the predominance of the $(1p)$ cavity mode for elongated rectangular mesas. 

When two mesas are biased to emit simultaneously, the far-field waves should be described in terms of the superposition of the $E$-fields generated from each mesa.  Thus, the total $E$-field depends on the phase difference between the macroscopic Josephson oscillations.  The observed pattern for A1$\parallel $A2 shown in Fig.~2(b) exhibits four-fold symmetry with respect to $\theta $, suggesting that the two mesas generate photons synchronously.  It is likely that $\psi $ and the helicity for A1$\parallel $A2 are dominated by the photon from A2, which emits more intensively than A1 (Fig.~2(c)).  Most importantly, we found that the axial ratio increases significantly from 2 for the individual emissions to 24 for the simultaneous emission.  We propose that this pronounced effect on the axial ratio is an indication of the phase synchronization between the two mesas as a result of EM coupling.

We observed the same behavior from other mesas shown in Fig.~1(a) and another supplemental sample, where each mesa showed slightly different polarization depending on the geometrical configuration~\cite{SM}.  The amplitude ratio $E_{0x} ⁄ E_{0y}$ explains the predominance of cavity resonances in the width direction over those in the length direction.  Since we injected the DC bias current into A1 and A2 using a left strip electrode as shown in Fig.~1(a), the resonance in the width direction may be degraded by a non-uniform current distribution in the mesa~\cite{Kakeya2012,Tsujimoto2014}.  This explanation is supported by the observation $E_{0x} / E_{0y} =1.4$ for A1 and, in contrast, 2.0 for supplemental mesas with a symmetrical electrode configuration~\cite{SM}. This technique allows for dynamic control of polarization by adjusting the current distribution in the mesa. 

In Fig.~3(a), we plot the polarization ellipses obtained by calculating the locus of $\bm{E}(t)$.  According to antenna theory for a transverse magnetic cavity, $E_{0x}$ (or $E_{0y}$) is proportional to the magnitude of the magnetic currents parallel to the $y$-axis (or $x$-axis).  Hence, in order to calculate $\bm{E}(t)$, we assume that the anisotropy is equal to the inverse mesa aspect, {\it i.e.}, $E_{0x} / E_{0y}=\ell /w$. This geometrical effect coincides with the numerical simulation for a locally heated square IJJ mesa~\cite{Asai2017}.  By comparing Fig.~3(a) with Fig.~2(c), we can estimate $\delta _{yx}$ to be $3 \pi / 4$ (135~deg) for A1 and $- \pi / 4$ ($-45$~deg) for A2.

Let us describe our results in terms of quantum mechanics.  The quantum-superposition state of the photon emitted from the parallel-biased mesa array of A1$\parallel $A2 is described as
\begin{equation}
\ket{\omega _{\textrm{A1}\parallel \textrm{A2}}, \bm{S}_{\textrm{A1}\parallel \textrm{A2}}}= 
\alpha \ket{\omega _{\textrm{A1}}, \bm{S}_{\textrm{A1}}} + 
\beta \ket{\omega _{\textrm{A2}}, \bm{S}_{\textrm{A2}}},  \notag
\end{equation} 
where $\bm{S}_i$ ($i=\textrm{A1, A2, or A1$\parallel $A2}$) represents the Stokes vector as a quantum number.  Two complex numbers $\alpha $ and $\beta $ represent the probability amplitudes ({\it i.e.}, $|\alpha |^2 + |\beta |^2 =1$) and phases of the unperturbed states $\ket{\omega _{\textrm{A1}}, \bm{S}_{\textrm{A1}}}$ and $\ket{\omega _{\textrm{A2}}, \bm{S}_{\textrm{A2}}}$, respectively.  According to the dispesrsion relation of the transverse JPW~\cite{Kadowaki1997}, angular frequencies $\omega _{\textrm{A1}}$ and $\omega _{\textrm{A2}}$ are determined by the wavenumbers $\bm{k}_{\textrm{A1}}$ and $\bm{k}_{\textrm{A2}}$ of the Josephson plasmons (quantized JPW) independently of the polarization.  Our concern is finding the 4$\times $4 perturbation matrix $V_m$ that satisfies
$
\begin{pmatrix}
\ket{\omega _{\textrm{A1}}',\bm{k}_{\textrm{A1}'}} \\
\ket{\omega _{\textrm{A2}}',\bm{k}_{\textrm{A2}'}} 
\end{pmatrix}
=V_{m}
\begin{pmatrix}
\ket{\omega _{\textrm{A1}},\bm{k}_{\textrm{A1}}} \\
\ket{\omega _{\textrm{A2}},\bm{k}_{\textrm{A2}}}
\end{pmatrix}.
$
Here, inter-mesa coupling $V_m$ causes perturbation and may include non-diagonal elements, which involve a general question in non-linear physics regarding the symmetry of the matrix.

Figure 3(b) shows polarization $\bm{S}_{\textrm{A1$\parallel $A2}}$ calculated by taking the superposition into consideration.  We used the actual intensity ratio obtained by measurement.  The orientation angle in the range of $0 < \psi < \pi / 2$ is in good agreement with Fig.~2(d).  It is noteworthy that the modulus $|\beta / \alpha |$ represents the degree of interaction between the two mesas and $|\beta / \alpha |=0.9$ coincides with our results.  Meanwhile, the argument of $\beta / \alpha $ corresponds to the phase difference between A1 and A2. 

Figure 4 shows the variation of axial ratio as a function of $\arg (\beta / \alpha )$.  As indicated by arrows, two singular states exhibiting perfect polarization are emitted when $|\beta / \alpha |<0.9$.  For example, two singular states with very large axial ratio can be observed at $\arg (\beta / \alpha )=45$~deg and at 135~deg when $|\beta / \alpha |=0.7$.  We suggest that such perfect polarization is attributed to coherent excitation of the Josephson plasmon.  See Supplemental Material for $|\beta / \alpha |$-dependence of $\bm{S}_{\textrm{A1$\parallel $A2}}$~\cite{SM}.

The origin of inter-mesa coupling $V_m$ arises from the propagation of JPWs through the Bi-2212 base crystal.  The inset of Fig.~4 shows a schematic cross-sectional view of the mesa array.  The dashed line in Fig.~4 represents the phase delay $2 \pi D/\lambda '$ due to the finite propagation time, where $D=58$~\textmu m is the interspace between the two mesas and $\lambda ' = \lambda /n$ is the effective wavelength.  We assume that JPWs can propagate from one mesa to another mesa through the base crystal and diffract at the mesa edge.  This situation strongly supports the view that the base crystal can mediate the EM interaction~\cite{Benseman2013b,Lin2013a}, the mechanism of which has been unclear in previous works.  Furthermore, we found that the total intensity of simultaneous emission $S_0$ reaches a maximal value as $2 \pi D/\lambda '$ coincides with a multiple of $\arg (\beta / \alpha )$~\cite{SM}.

In conclusion, we demonstrated the synchronization of macroscopic Josephson oscillations in two simultaneously biased Bi-2212 IJJ mesas coupled via a base crystal by measuring the complete Stokes parameters.  We used an achromatic QWP and a linear WGP to analyze the orthogonal components of the emitted $E$-fields.  We proved that the coherent radiation is elliptically polarized with the major axis oriented in the width direction.  Most importantly, we observed a significant increase in the axial ratio for simultaneous emission, suggesting that Josephson plasma in the Bi-2212 base crystal can mediate an interaction between two individual mesas.  This finding represents a possible means of manipulating the synchronization of IJJ arrays, and is the most promising way to increase the integrated radiation power.

\section*{Acknowledgment}
This work was supported by the Japan Society for the Promotion of Science (JSPS) KAKENHI (Grant No.~26286006, No.~15KK0204, and No.~19H02540), JSPS-Centre national de la recherche scientifique (CNRS) Bilateral Program (Grant No.~120192908), and the Program to Disseminate the Tenure Tracking System at the University of Tsukuba.  The Bi-2221 single crystal was provided by Y.~Nakagawa at Kyoto University. The authors thank H.~Asai, S.~Kawabata, M.~Machida, and T.~Koyama for their valuable discussions.


\bibliography{MyCollection}

\clearpage
\begin{figure}[p] 
	\includegraphics[width=0.85\textwidth ,clip]{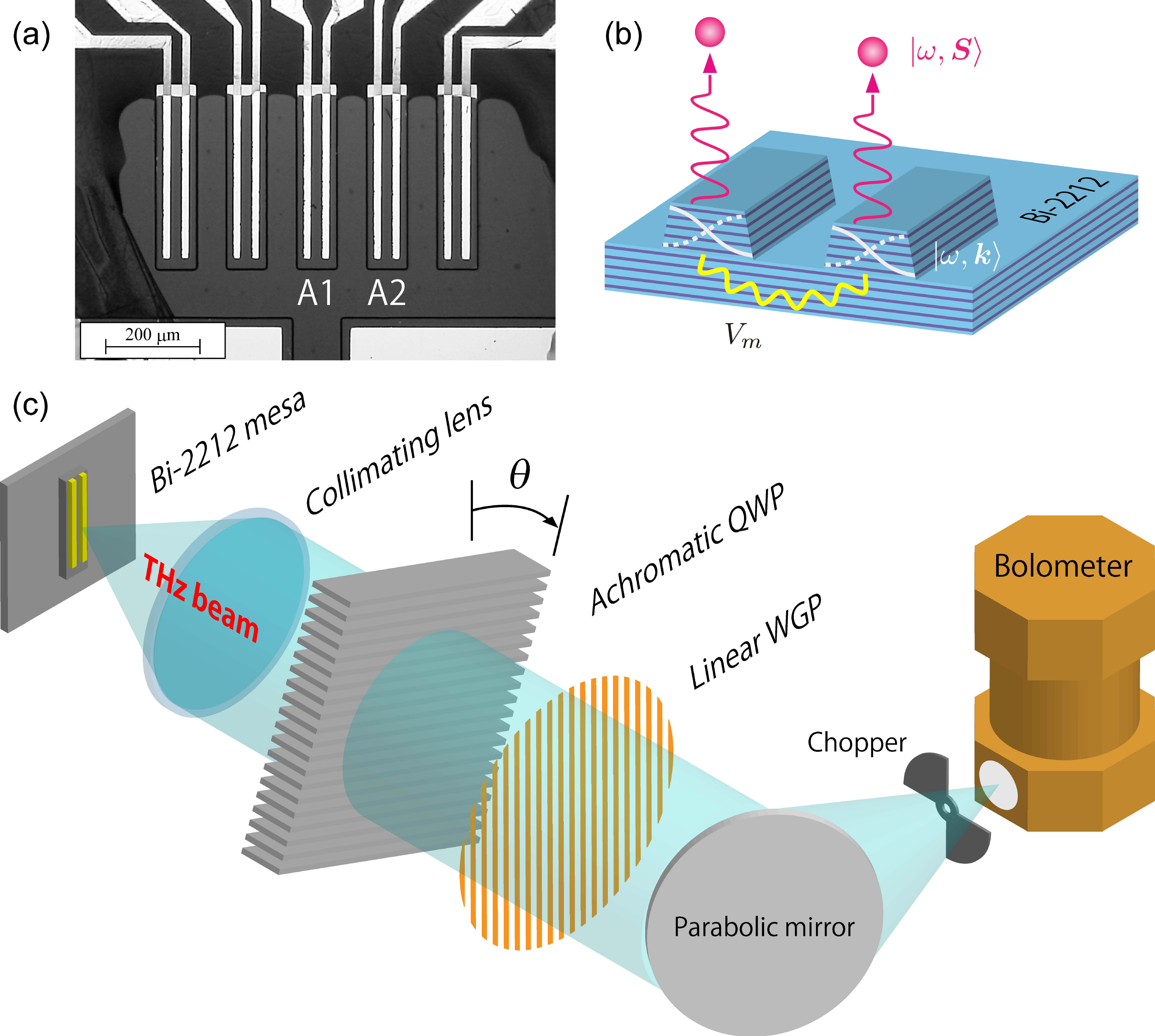} \\
	\caption{
(a) Optical microscopy image of the Bi-2212 mesa array.  (b) Schematic view of the synchronized array.  (c) Schematic of the optical configuration for Stokes parameter measurement. 
}
\end{figure}%

\clearpage
\begin{figure}[p] 
	\includegraphics[width=0.85\textwidth ,clip]{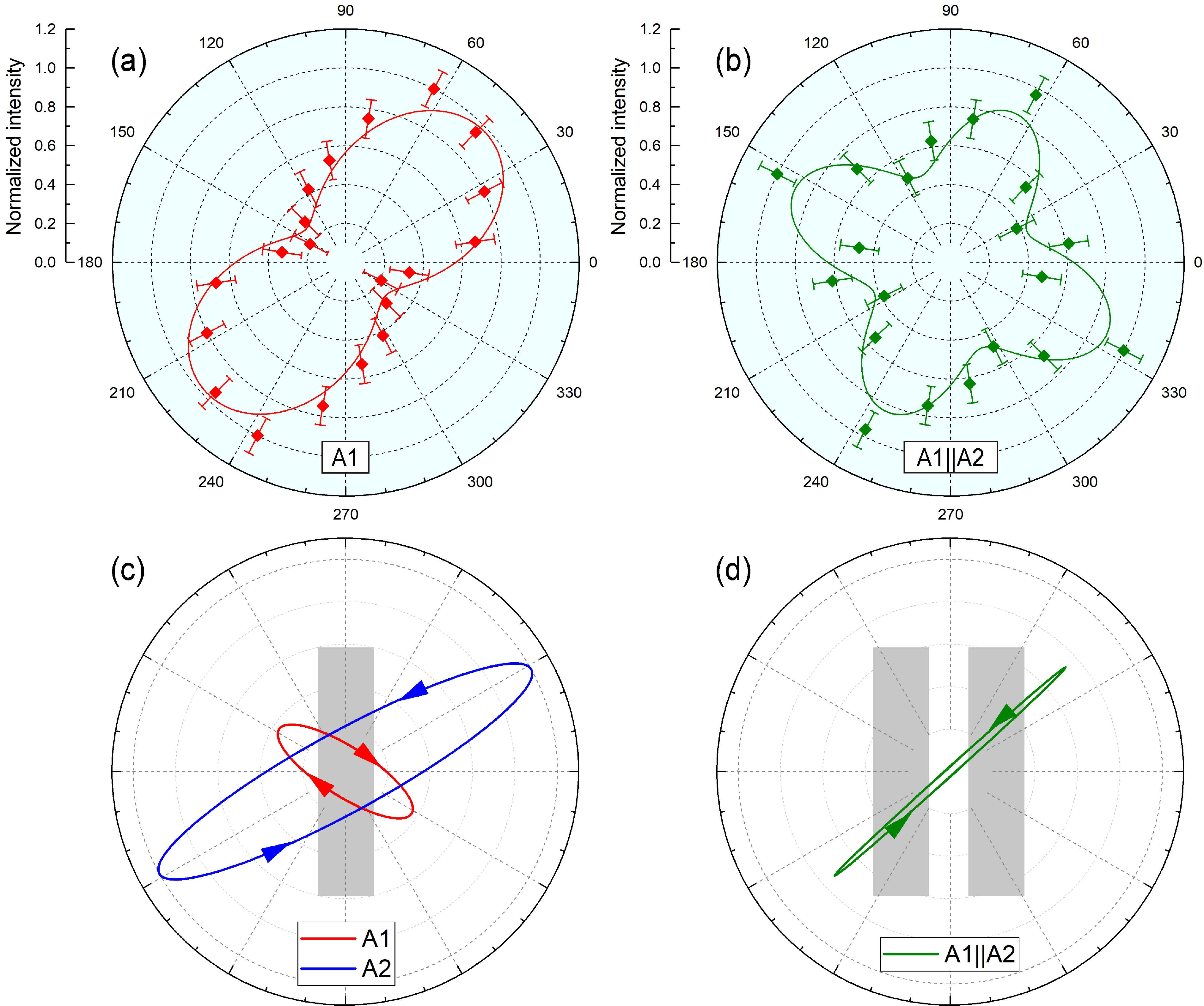} \\
	\caption{
Polar plots of the maximum bolometer output as a function of QWP angle $\theta $ for (a) A1 and A2, and (b) A1$\parallel $A2.  The solid lines represent results calculated using the Stokes parameters.  Polarization ellipses for (c) A1 (red) and A2 (blue), and (d) A1$\parallel $A2.
}
\end{figure}%

\clearpage
\begin{figure}[p] 
	\includegraphics[width=0.85\textwidth ,clip]{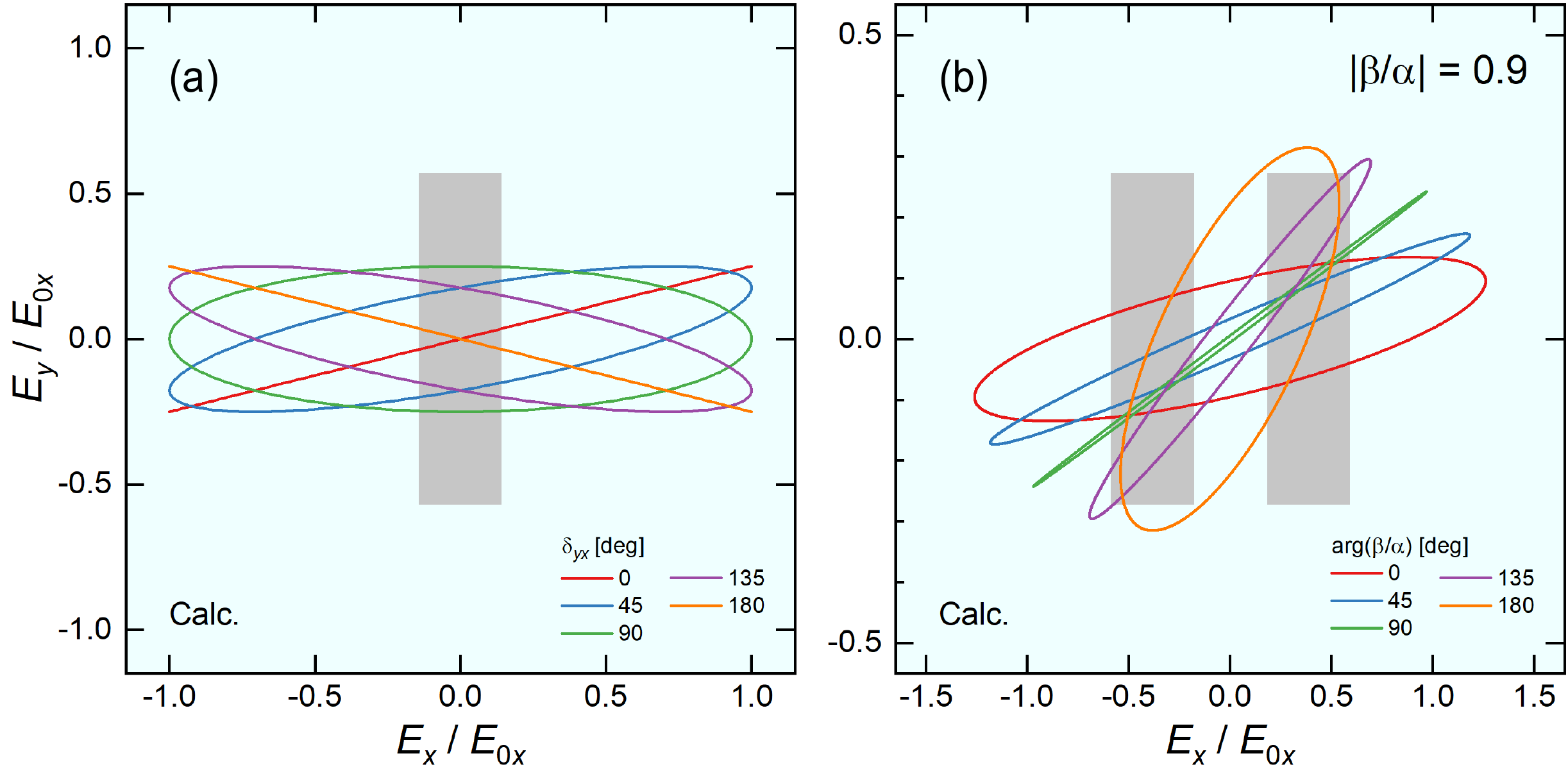} \\
	\caption{
Calculation of the polarization ellipse for (a) a single mesa and (b) two arrayed mesas. 
}
\end{figure}%

\clearpage
\begin{figure}[p] 
	\includegraphics[width=0.85\textwidth ,clip]{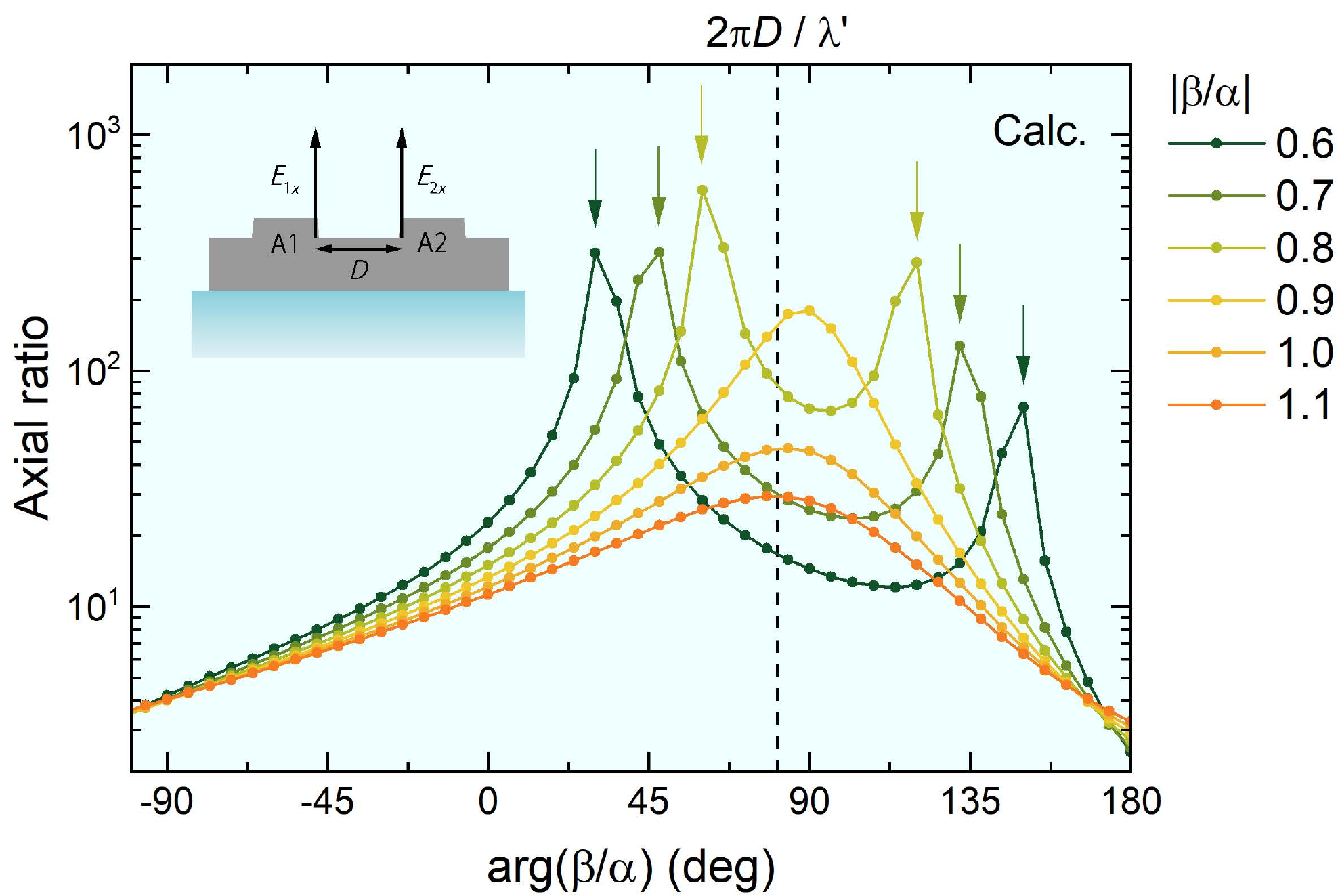} \\
	\caption{
Calculation of the axial ratio as a function of $\arg (\beta / \alpha )$. Inset: schematic side view of the Bi-2212 mesa array.
}
\end{figure}%

\end{document}


\pagenumbering{roman}


\title{Supplemental Material:\\
Mutually synchronized macroscopic Josephson oscillations demonstrated by polarization analysis of superconducting terahertz emitters}

\author{M. Tsujimoto$^{1,2}$}%
 \altaffiliation[Corresponding author: ]{tsujimoto@ims.tsukuba.ac.jp}

\author{S. Fujita$^3$}
\author{G. Kuwano$^2$}
\author{K. Maeda$^3$}
\author{A. Elarabi$^3$}
 \altaffiliation[Present address: ]{Okinawa Institute of Science and Technology Graduate University, Onna, Okinawa 904-0495, Japan}
\author{J. Hawecker$^4$}
\author{J. Tignon$^4$}
\author{J. Mangeney$^4$}
\author{S. S. Dhillon$^4$} 
\author{I. Kakeya$^3$}
 \altaffiliation[Corresponding author: ]{kakeya@kuee.kyoto-u.ac.jp}

\affiliation{%
$^1$Faculty of Pure and Applied Sciences, University of Tsukuba, 1-1-1 Tennodai, Tsukuba, Ibaraki 305-8573, Japan
}%

\affiliation{%
$^2$Graduate School of Pure and Applied Sciences, University of Tsukuba, 1-1-1 Tennodai, Tsukuba, Ibaraki 305-8573, Japan
}%

\affiliation{%
$^3$Department of Electronic Science and Engineering, Kyoto University, Nishikyo-ku, Kyoto 615-8510, Japan
}%

\affiliation{%
$^4$Laboratoire de Physique de l’Ecole normale sup\'{e}rieure, ENS, Universit\'{e} PSL, CNRS, Sorbonne Universit\'{e}, Universit\'{e} de Paris, F-75005 Paris, France
}%

\date{\today}

\maketitle


\section{Verification of quarter-wave plate device}
In order to measure the Stokes parameters, we used a lab constructed quarter-wave plate (QWP) consisting of a stack of parallel metal-plate waveguides.  The thickness of the spacer was set to 2.0~mm so that the bandwidth of the QWP coincides with the frequency range of our superconducting terahertz source. 

To characterize the QWP, we performed terahertz time-domain spectroscopy (THz-TDS) using an optical system based on a comb-type photoconductive antenna and ZnTe crystal electro-optic detection with an 800-nm femtosecond pump laser~\cite{Madeo2010,Maussang2016}.  In this setup, the QWP and a pair of linear polarizers were placed in the optical path according to~\cite{Nagai2015}.  The measurement results shown in Fig.~S1(a) prove that the transmitted $E$-field component parallel to the plate was delayed by $\pi /2$ in the frequency range of 0.3–-0.8~THz. Figure S1(b) shows the trajectory of the $E$-field vector composed of orthogonal time-domain signals.  These results demonstrate that the QWP device functions as a $\pi /2$ retarder for parallel transmission.

\section{Formulation of the Stokes parameter analysis}
A useful visual representation of the polarized wave is given by 
\begin{equation}
\frac{E_{x}^2(z,t)}{E^2_{0x}}+\frac{E_{y}^2(z,t)}{E^2_{0y}}  
- \frac{2E_{x}(z,t)E_{y}(z,t)}{E_{0x}E_{0y}}\cos \delta _{yx}=
\sin ^2 \delta _{yx} 
\end{equation}
where $\delta _{yx} = \delta _{y} - \delta _{x}$ is the lateral phase difference.  Equation (1) describes the elliptical polarization when $E_{x} \neq E_{y}$ and $\delta _{yx} \neq 0$.  Figure S2 shows the polarization ellipse that defines the optical parameters.  We define the $x$- and $y$-axes as the directions parallel to the Bi$_2$Sr$_2$CaCu$_2$O$_{8+\delta }$ (Bi-2212) mesa width and length, respectively.  The EM wave propagates along the $z$-direction.

By taking the time average, Eq.~(1) can be transformed to the intensity domain~\cite{Collett1968}, 
\begin{equation}
S^2_{0} = S^2_{1} + S^2_{2} + S^2_{3}, 
\end{equation}
where the set of four independent polarization parameters, 
\begin{equation}
\bm{S}=
\begin{pmatrix}
S_0 \\
S_1 \\
S_2 \\
S_3
\end{pmatrix}
=
\begin{pmatrix}
E^2_{0x} + E^2_{0y} \\
E^2_{0x} - E^2_{0y} \\
2E_{0x}E_{0y} \cos \delta \\
2E_{0x}E_{0y} \sin \delta 
\end{pmatrix},
\end{equation}
are called the Stokes parameters~\cite{Stokes1851}.  These parameters are widely used to describe polarization primarily because the polarization ellipse is not directly accessible by measurement~\cite{Wolf1954}.  The parameter $S_0$ represents the total intensity of the optical field, $S_1$ represents the predominance of linearly horizontally polarized light over linearly vertically polarized light, $S_2$ represents the predominance of linear $+45$~deg polarized light over linear $-45$~deg polarized light, and the fourth parameter $S_3$ represents the predominance of right circular polarization over left circular polarization.  The advantage of the Stokes parameters is that they can be measured because they are represented in terms of the intensities of specific polarizations.

The polarization ellipse shown in Fig.~S2 can be described in terms of two angles: the orientation angle $\psi $ and ellipticity angle $\chi $.  These angles can be determined from the Stokes parameters~\cite{Born1999} as
\begin{equation}
\psi = \frac{1}{2} \tan ^{-1} \left( \frac{S_2}{S_1} \right) 
\quad 
(0 < \psi < \pi)
\end{equation}
\begin{equation}
\chi = \frac{1}{2} \sin ^{-1} \left( \frac{S_3}{S_0} \right) 
\quad 
(-\frac{\pi }{4} < \chi < \frac{\pi }{4}). 
\end{equation}
The major axis $\xi $ is directed along an axis rotated through $\psi $, and the axial ratio equals $1/ \tan \chi $.

The intensity measured using the bolometer as a function of the QWP angle $\theta $ is given by~\cite{Collett1992} 
\begin{equation}
I(\theta )=\frac{1}{2} (S_{0} + S_{1} \cos ^2 2\theta  
+ S_{2} \cos 2\theta \sin 2\theta + S_{3} \sin 2\theta ).
\end{equation}
The squared and product terms can be rewritten using the trigonometric half-angle formula to yield 
\begin{equation}
I(\theta )=\frac{1}{2} (A + B \sin 2\theta + C \cos 4\theta + D \sin 4\theta ),
\end{equation}
where
\begin{align}
A=S_{0} 
& \quad &
 B=S_{3} 
& \quad &
  C=\frac{S_{1}}{2} 
& \quad &
   D=\frac{S_{2}}{2}. 
\end{align}
Equation (7) is a truncated Fourier series consisting of a DC term, a second harmonic term, and two fourth harmonics terms.  Thus, the four Fourier coefficients in Eq.~(8) can be determined by the following equations:
\begin{equation}
\begin{split}
&A=\frac{2}{N_m} \sum_{n=1}^N I_{n} \\
&B=\frac{4}{N_m} \sum_{n=1}^N I_{n} \sin 2\theta _{n} \\
&C=\frac{4}{N_m} \sum_{n=1}^N I_{n} \cos 4\theta _{n} \\
&D=\frac{4}{N_m} \sum_{n=1}^N I_{n} \sin 4\theta _{n}
\end{split}
\end{equation}
where $I_{n}$ is the intensity measured at angle $\theta _{n}$ and $N_m$ is the number of measurements.  From Eq.~(8), the Stokes parameters are found to be 
\begin{equation}
\begin{split}
&S_{0}=A-C 
 \qquad 
S_{1}=2C \\
&S_{2}=2D 
 \qquad 
S_{3}=B. 
\end{split}
\end{equation}
We measured $N_{m}=10$ data every 18~deg in the range of $0 \leq \theta <180$ to calculate the Stokes parameters using Eqs.~(8), (9), and (10).

\section{Emission characteristics}
Figure S3(a) shows the current-voltage characteristics (IVCs) of the Bi-2212 mesas.  The red and blue curves represent the data obtained from mesas A1 and A2, respectively, and the green curve is the IVC from the two mesas connected in parallel and biased simultaneously using a single voltage source ({\it i.e.}, A1$\parallel $A2).  The bath temperature was fixed at $T_{b}= 15.0$~K.  Each IVC curve shows the large hysteresis typical of underdamped IJJs.  We used a silicon-composite bolometer (Infrared Laboratories, Inc.) and lock-in amplifier with a chopper to detect terahertz radiation.  Figure S3(b) shows the bolometer output as a function of applied current.  EM radiation was observed in the resistive state, with maximum outputs at 5.4~mA for A1 and 6.0~mA for A2 as indicated by the arrows in Figs.~S3(a) and S3(b). 

As shown in Fig.~S3(b), the maximum output for A1 was approximately 13\% of that for A2.  This difference arises from in-plane non-uniformity of the superconducting order parameters in the Bi-2212 base crystal.  The inset of Fig.~S3(b) shows the temperature dependence of the $c$-axis resistance.  The onset $T_c$ for A1 and A2 were 77.7~K and 81.4~K, respectively. 

Global emission occurs when A1 and A2 are connected in parallel and biased on the outermost IVC branch (see the green line in Fig.~S3(b)).  The maximum intensity is obtained at 16.8~mA, which is higher than the sum of the bias currents (11.4~mA) for the maximum emission powers of A1 and A2 individually. Furthermore, the emission voltage was considerably lower than those of A1 and A2.  As indicated by the arrows in Fig.~S3(a), the emission voltage was 0.53~V for A1$\parallel $A2, whereas that for A1 and A2 was 0.74~V.  This can be partly explained by considering the local temperature increase, which is more pronounced for mesa arrays~\cite{Benseman2013b} and allows for a reduction in the emission voltage.  The effective local temperature near the biased mesas increases considerably owing to significant Joule heating, especially in the high-bias regime~\cite{Wang2009,Wang2010,Gross2012,Benseman2013c,Minami2014,Tsujimoto2014,Benseman2015}.

A lab-constructed Fourier transform interferometer based on split mirrors was used to measure the radiation frequencies.  For the configuration employed the frequency resolution was 10~GHz.  Figure S4(a) shows the I-V characteristic curve for A2, while Fig.~S4(b) shows the emission spectra obtained at 6.5~mA and 9.9~mA.  According to the AC Josephson effect~\cite{Josephson1962}, application of a DC voltage $V$ leads to an AC current and electromagnetic emission at the Josephson frequency in the form of $f_{J}=2eV ⁄ h N$, where $e$ is the elementary charge, $h$ is the Planck constant, and $N$ is the number of junctions contributing to the emission.  By substituting the radiation frequencies and voltages into the above equation, we can estimate that at most $N \sim 600$--700 junctions contribute to emission.  The small steps in the outermost I-V branches indicate that emission takes place in the inner regions of multiple I-V branches~\cite{Tsujimoto2012a}, where the number of resistive junctions is reduced from the total number. 

Figure S5 shows the $T_b$ dependence of the I-V characteristics with color-coded bolometer outputs.  The results demonstrate stable emission on the outermost branches in the low-bias regime at temperatures below $T_{b}=55.0$~K.

\section{Supplemental data from a mesa array of differing design}
We present supplemental data obtained from a mesa array of differing design.  The size of the mesas were almost identical at 77$\times $350$\times $1.2~\textmu m$^3$ with an interspace of 78~\textmu m.  Here, the width was measured at the mesa bottom, and the width at the top was 72~\textmu m. 

Figure S6 displays polar plots of the maximum bolometer output for mesa (a) B1 and (b) B2 at $T_{b}= 42.0$~K.  Figure S6(c) shows the same plot for global emission from mesa array (B1$\parallel $B2).  Data is plotted in the same way as was done in Fig.~2 of the main text. 

The obtained Stokes parameters are summarized in Table SI.  Figure S6(d) shows the polarization ellipses.  In the case of B1$\parallel $B2, we observed a significant increase in axial ratio from $\sim $8 to 20.  This behavior is consistent with A1$\parallel $A2, and the other polarization parameters were similar to those for the main sample.  Note that we did not examine the possibility of synchronization with more than two mesas.

\section{Calculation of polarization for various coupling strengths}
Figure S7(a–f) shows various polarization characteristics $\bm{S}_{\textrm{A1$\parallel $A2}}$ plotted for different $|\beta / \alpha |$ values.  The modulus $|\beta / \alpha |$ is the degree of interaction between the two mesas, $|\beta / \alpha |=0.9$ coinciding with the experimental results.  Meanwhile, $\arg (\beta / \alpha )$ corresponds to the phase difference between the two mesas. 

As indicated by arrows in Fig.~4 in the main text, two singular states exhibiting perfect linear polarization are emitted when $|\beta / \alpha |<0.9$.  For example, two singular states with very large axial ratios can be clearly observed in Fig.~S7(b) at $\arg (\beta / \alpha )=45$~deg and at 135~deg. 

\newpage

\section{Calculation of the total intensity as a function of inter-mesa space}
We simulated the total intensity of the global radiation emitted from the mesa arrays.  The total intensity was derived from the first Stokes parameter $S_0$ and was measured using a bolometer in our experiment.

The orthogonal components for each mesa are given as follows: 
\begin{equation}
\begin{split}
&E_{x1}(t)=E_{0x} \cos (\omega t) \\
&E_{y1}(t)=E_{0y} \cos (\omega t + \delta _{yx1}) \\
&E_{x2}(t)=E_{0x} \cos (\omega t + \delta _{21}) \\
&E_{y2}(t)=E_{0y} \cos (\omega t + \delta _{21} + \delta _{yx2}), 
\end{split}
\end{equation}
where 1 and 2 denote mesa A1 and A2, respectively.  The relative phase $\delta _{21}$ represents the phase difference between the $E_x$ of each mesa.  The linear combination of these fields allows for calculation of the total intensity $S_0$.  Using the sum formula for the cosine function, $\cos \phi + \cos (\phi + \gamma ) = 
\frac{\sin \gamma }{\sin (\delta _{21}/2)}
\cos (\phi + \frac{\gamma }{2})$, we can obtain the total intensity as
\begin{equation}
S_{0}=
\left\{
E_{0x}
\frac{\sin \delta _{21}}{\sin (\delta _{21}/2)}
\right\}^2
+ \left\{
E_{0y}
\frac{\sin (\delta _{21}+\delta _{yx2}-\delta _{yx1})}{\sin (\frac{\delta _{21}+\delta _{yx2}-\delta _{yx1}}{2})}
\right\}^2 .
\end{equation}
In Fig.~S8, we plot $S_0$ for various $\delta _{yx2}-\delta _{yx1}$ as a function of $D$.  To calculate $S_0$, we use $E_{0x} / E_{0y} = \ell / w$ and $D/\lambda '=\delta _{21}/2\pi $.  The results clearly indicate that the total intensity has local maxima when $D$ coincides with an integer multiple of $\lambda '$ independently of $\delta _{yx}$.  When synchronized, the total intensity depends strongly on $D$, whereas its peak value is only marginally affected by $\delta _{yx}$.  This allows for the possibility of increasing the total intensity and controlling the consequent polarization by adjusting $D$.


\bibliography{MyCollection}

\clearpage
\begin{figure}[p] 
	\includegraphics[width=0.85\textwidth ,clip]{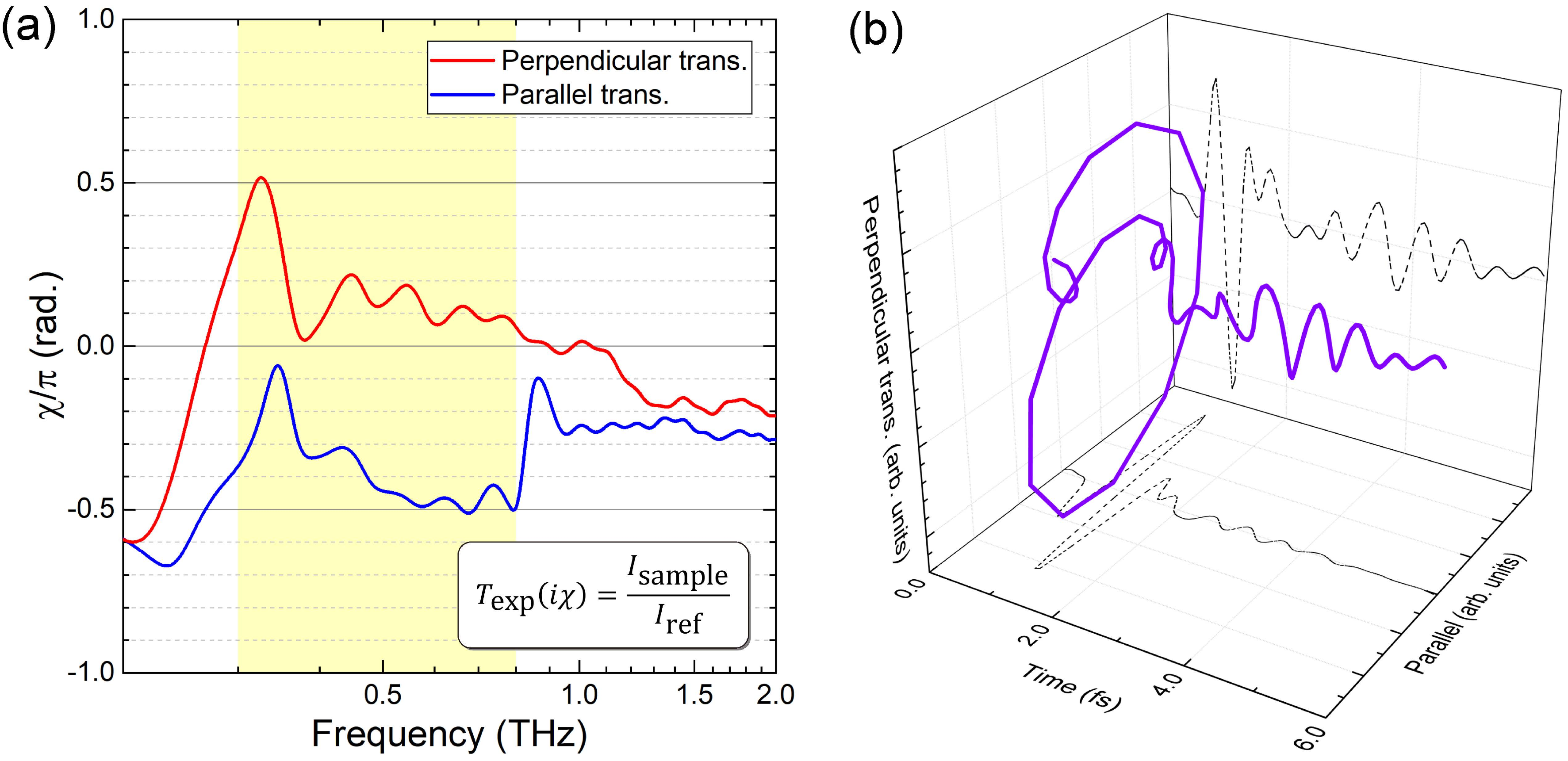} \\
	\caption{
Verification of the fabricated QWP device using the THz-TDS setup.  (a) Phase changes due to the QWP device for $E$-field components perpendicular (red) and parallel (green) to the metal-plate waveguides.  (b) Temporal evolution of $E$-field transmitted through the QWP device.
}
\end{figure}%

\clearpage
\begin{figure}[p] 
	\includegraphics[width=0.85\textwidth ,clip]{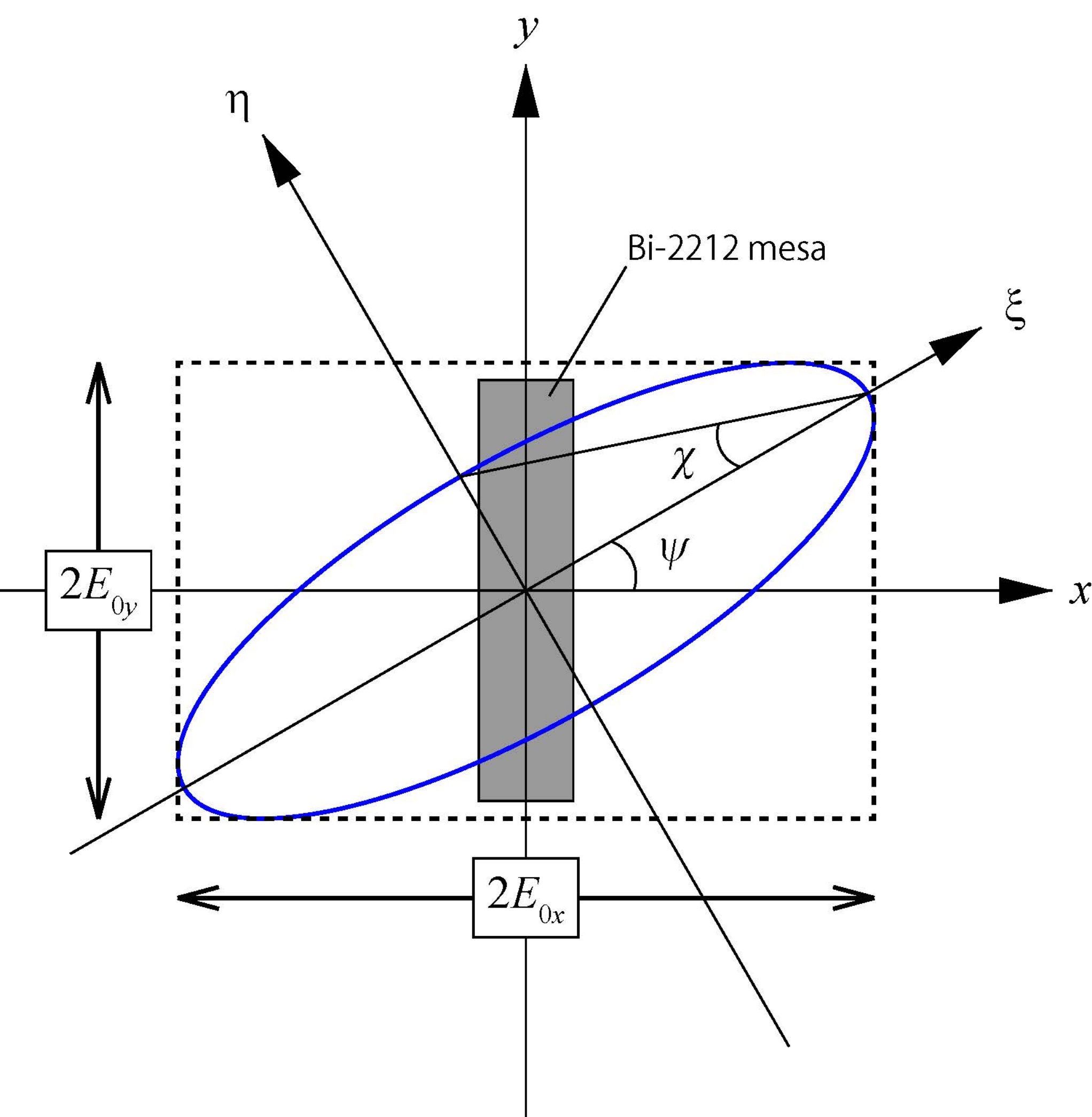} \\
	\caption{
Polarization ellipse defining the optical parameters. 
}
\end{figure}%

\clearpage
\begin{figure}[p] 
	\includegraphics[width=0.85\textwidth ,clip]{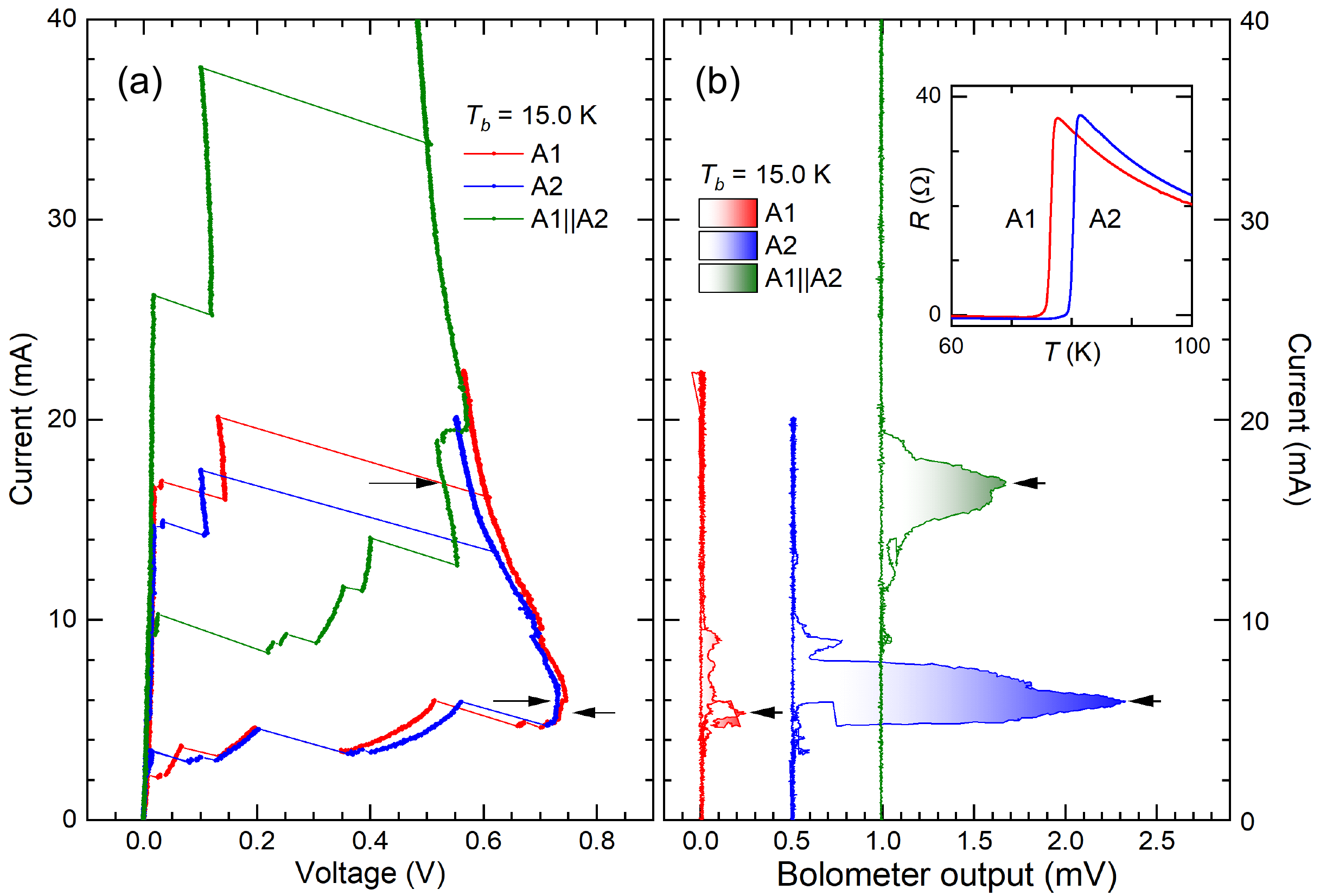} \\
	\caption{
(a) Current-voltage characteristics and (b) lock-in signals from the silicon-composite bolometric terahertz detector for mesa A1 (red), mesa A2 (blue), and A1$\parallel $A2 (parallel bias configuration, green). The bath temperature was fixed at $T_{b} = 15.0$~K.  Inset of (b): temperature dependence of the $c$-axis resistance.
}
\end{figure}%

\clearpage
\begin{figure}[p] 
	\includegraphics[width=0.85\textwidth ,clip]{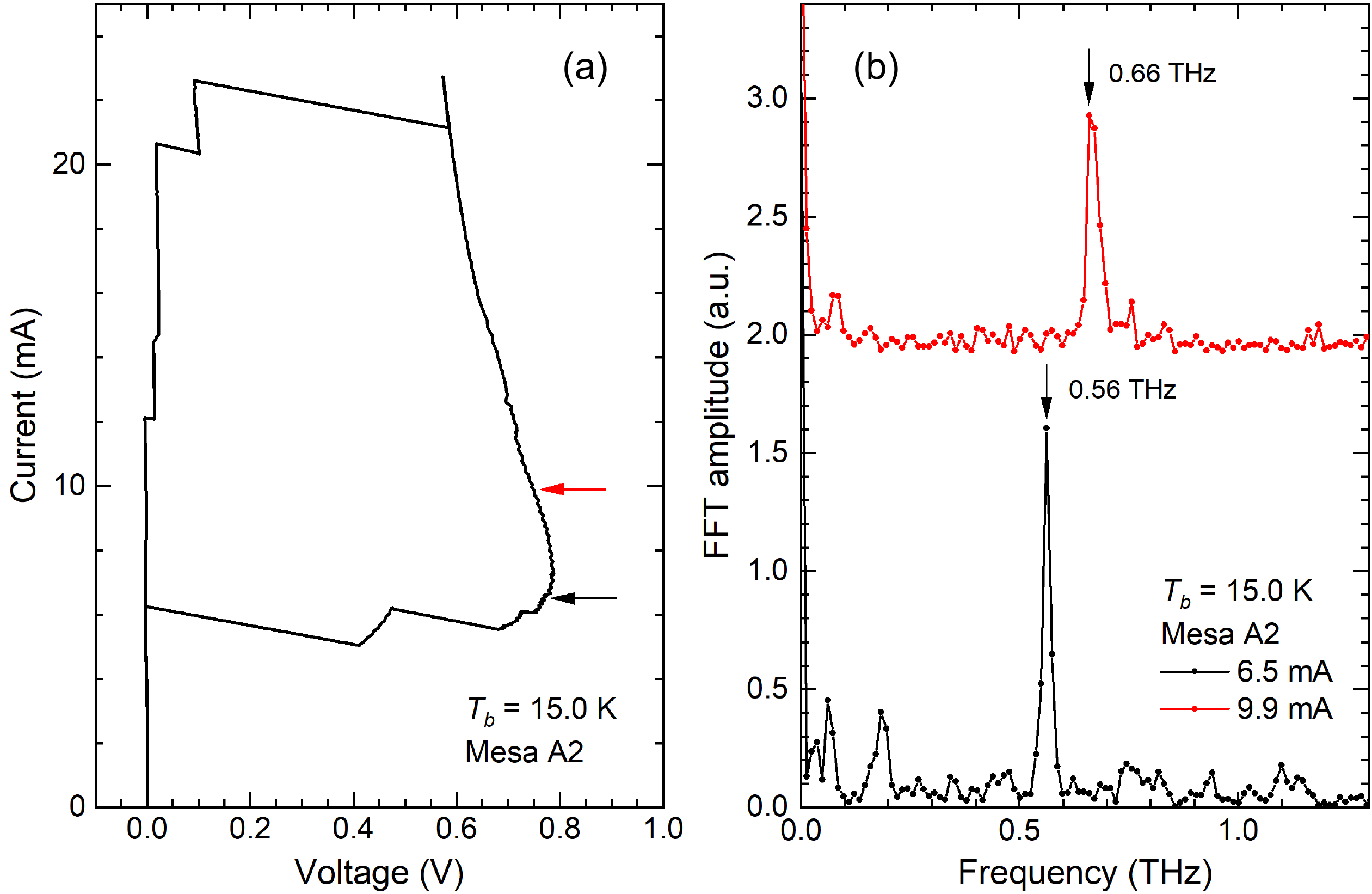} \\
	\caption{
(a) Current-voltage characteristics for mesa A2 measured at $T_{b} = 15.0$~K.  (b) Emission spectra obtained using the Fourier transform interferometer.  The bias conditions are indicated by arrows in (a). 
}
\end{figure}%

\clearpage
\begin{figure}[p] 
	\includegraphics[width=0.85\textwidth ,clip]{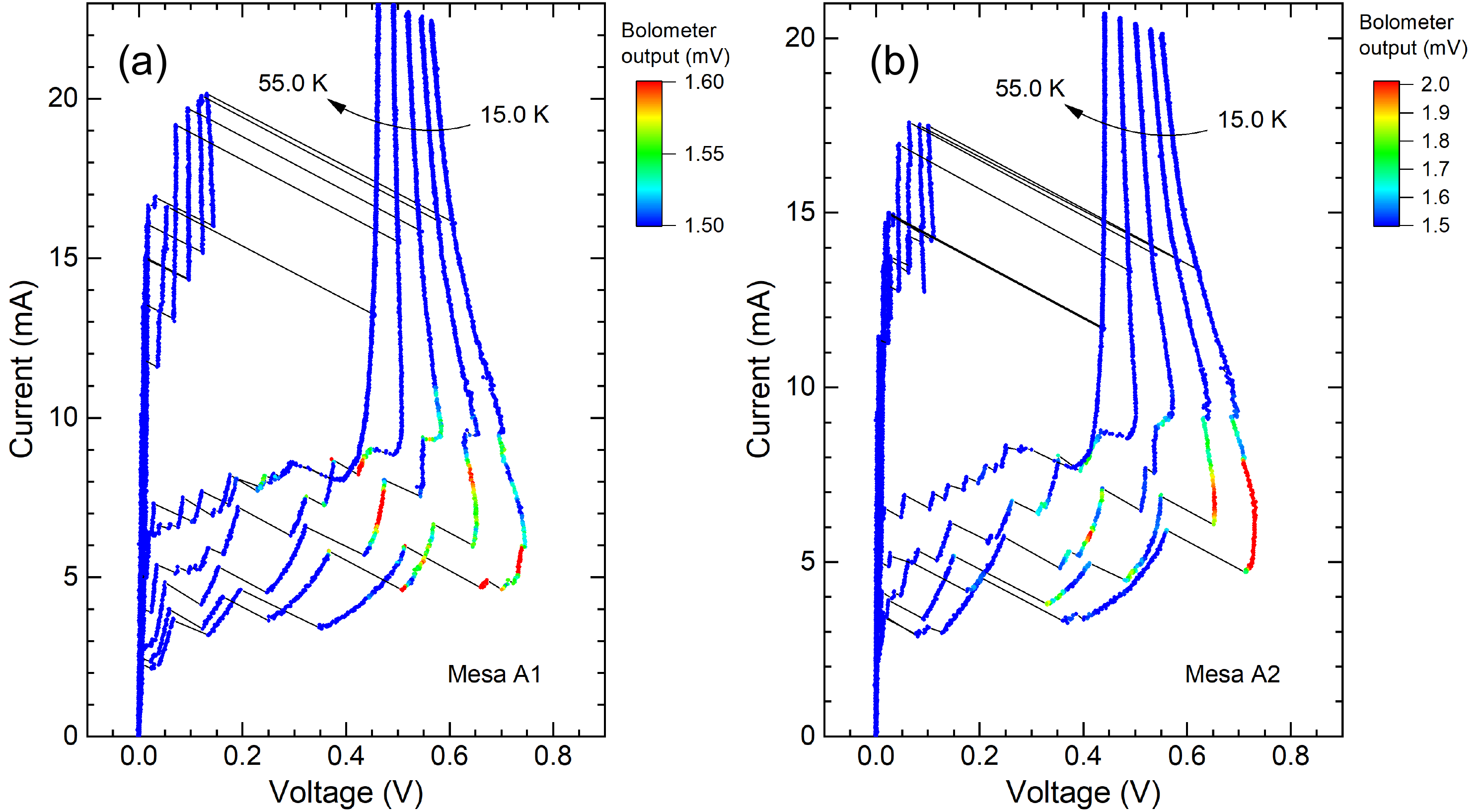} \\
	\caption{
$T_b$ dependence of the current-voltage characteristics plotted with color-coded bolometer outputs for mesas (a) A1 and (b) A2.
}
\end{figure}%

\clearpage
\begin{figure}[p] 
	\includegraphics[width=0.85\textwidth ,clip]{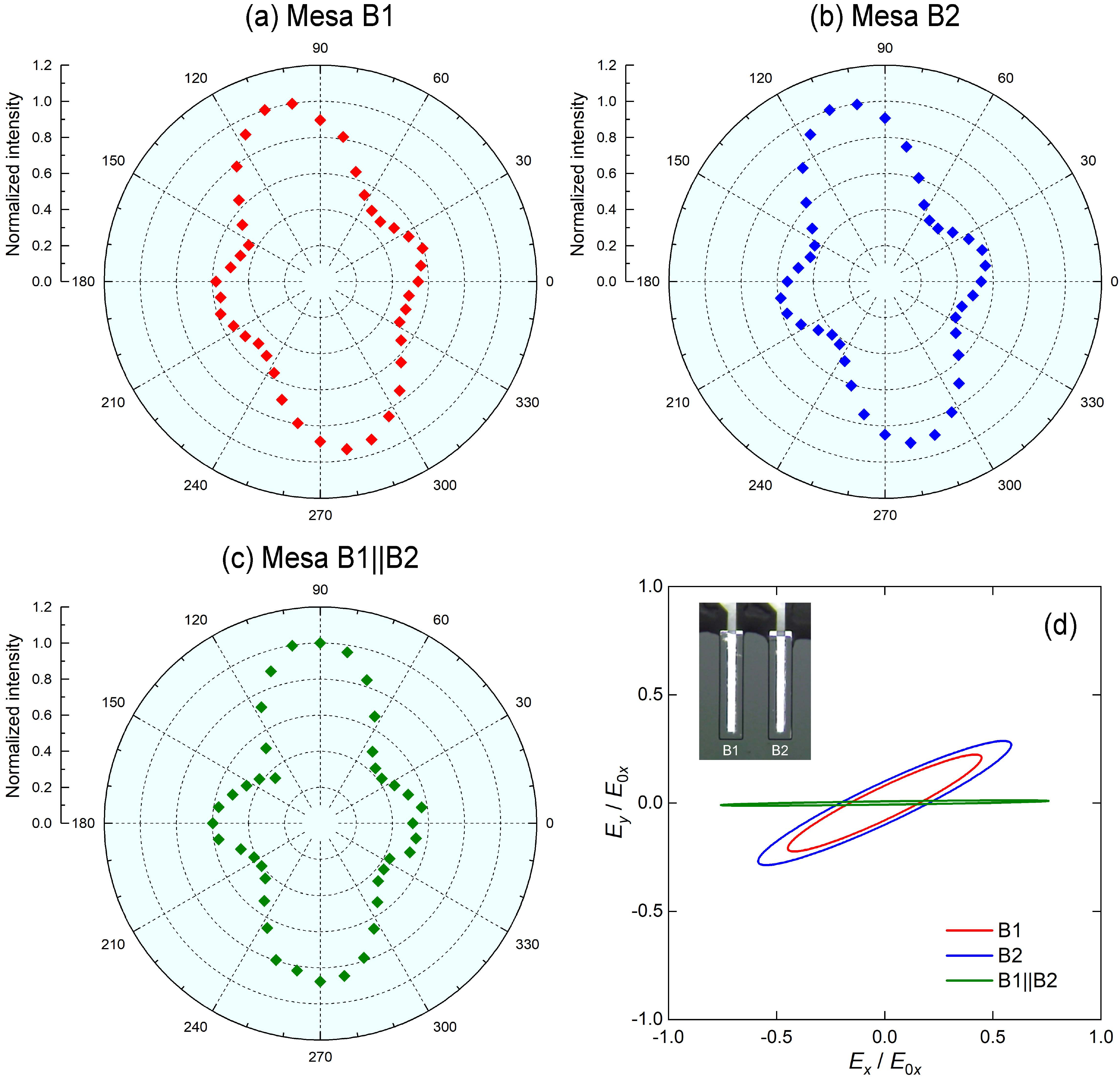} \\
	\caption{
Supplemental data obtained from a mesa array of a differing design.  Polar plots of the maximum bolometer output as a function of QWP angle $\theta $ for mesa (a) B1, (b) B2, and (c) B1$\parallel $B2 (parallel connection).  (d) Polarization ellipses.  Inset: Optical microscopy image of the sample.
}
\end{figure}%

\clearpage
\begin{table}[p]
\caption{
Polarization parameters for two mesa arrays: Stokes parameters $S_0$, $S_1$, $S_2$, and $S_0$, in-plane phase difference $\delta _{yx}$, and axial ratio.
}
\begin{tabular}{cccccccc}
\toprule
Mesa & A1 & A2 & A1$\parallel $A2 & \hspace{1em} & B1 & B2 & B1$\parallel $B2 \\ \midrule
$S_0$   & 1.00  & 1.00  & 1.00  & \hspace{1em} & 1.00  & 1.00  & 1.00   \\
$S_1$   & -0.09 & -0.24 & -0.06 & \hspace{1em} & 0.30  & 0.40  & 0.40   \\
$S_2$   & 0.06  & -0.38 & -0.61 & \hspace{1em} & 0.37  & 0.48  & 0.48   \\
$S_3$   & 0.57  & -0.36 & -0.05 & \hspace{1em} & -0.14 & -0.18 & -0.18  \\
$\delta _{yx}$~(deg)    & 95.9  & -44.0 & -4.6  & \hspace{1em} & -21.0 & -20.5 & -37.5 \\
Axial ratio   & 1.4   & 2.6   & 24.0  & \hspace{1em} & 8.4   & 8.5   & 20.0   \\ \bottomrule
\end{tabular}
\end{table}

\clearpage
\begin{figure}[p] 
	\includegraphics[width=0.85\textwidth ,clip]{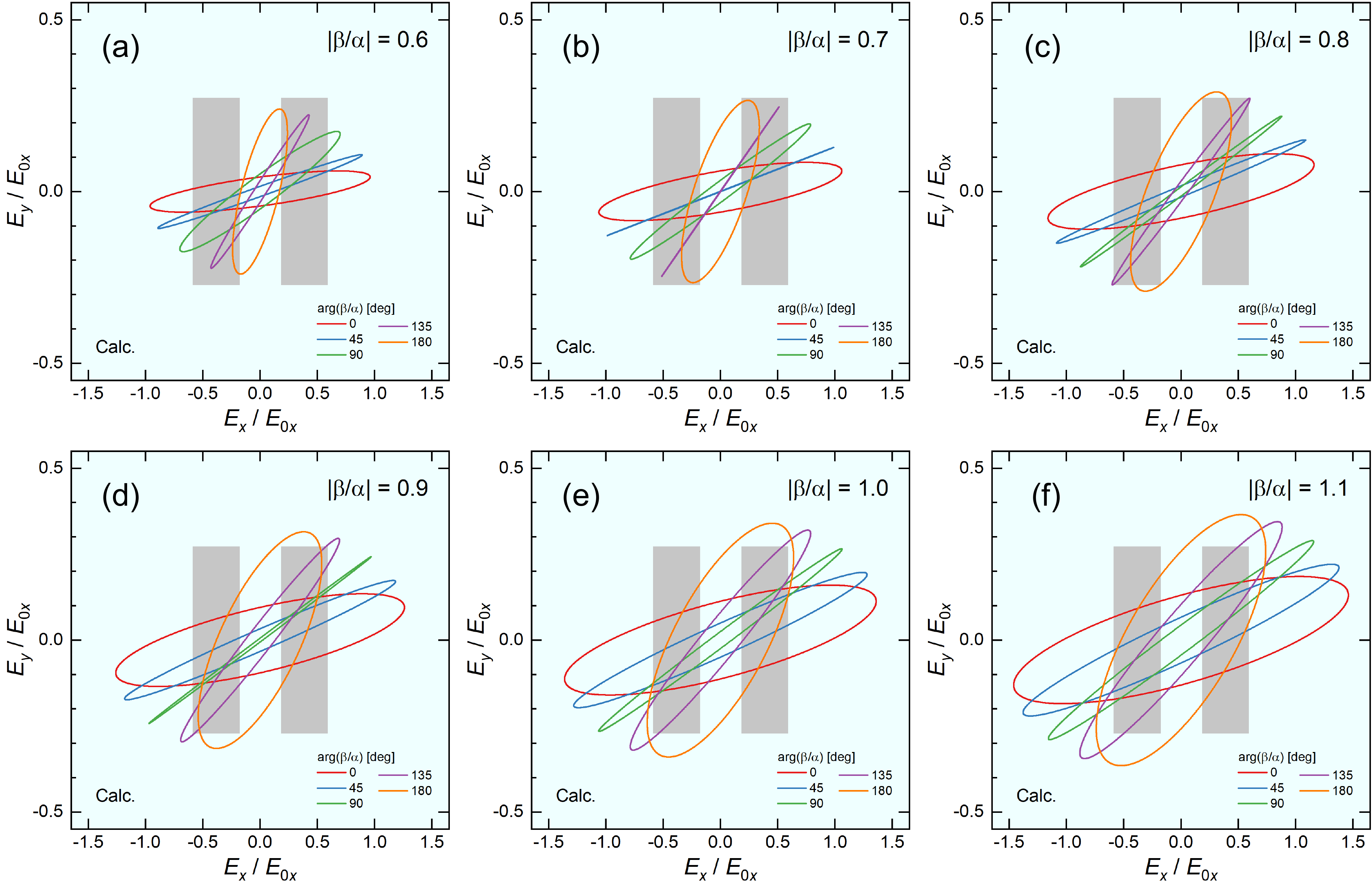} \\
	\caption{
Calculation of polarization $S_\textrm{A1$\parallel $A2}$ for different $|\beta / \alpha |$ values:  (a) $|\beta / \alpha |=0.6$, (b) 0.7, (c) 0.8, (d) 0.9, (e) 1.0, and (f) 1.1. 
}
\end{figure}%

\clearpage
\begin{figure}[p] 
	\includegraphics[width=0.85\textwidth ,clip]{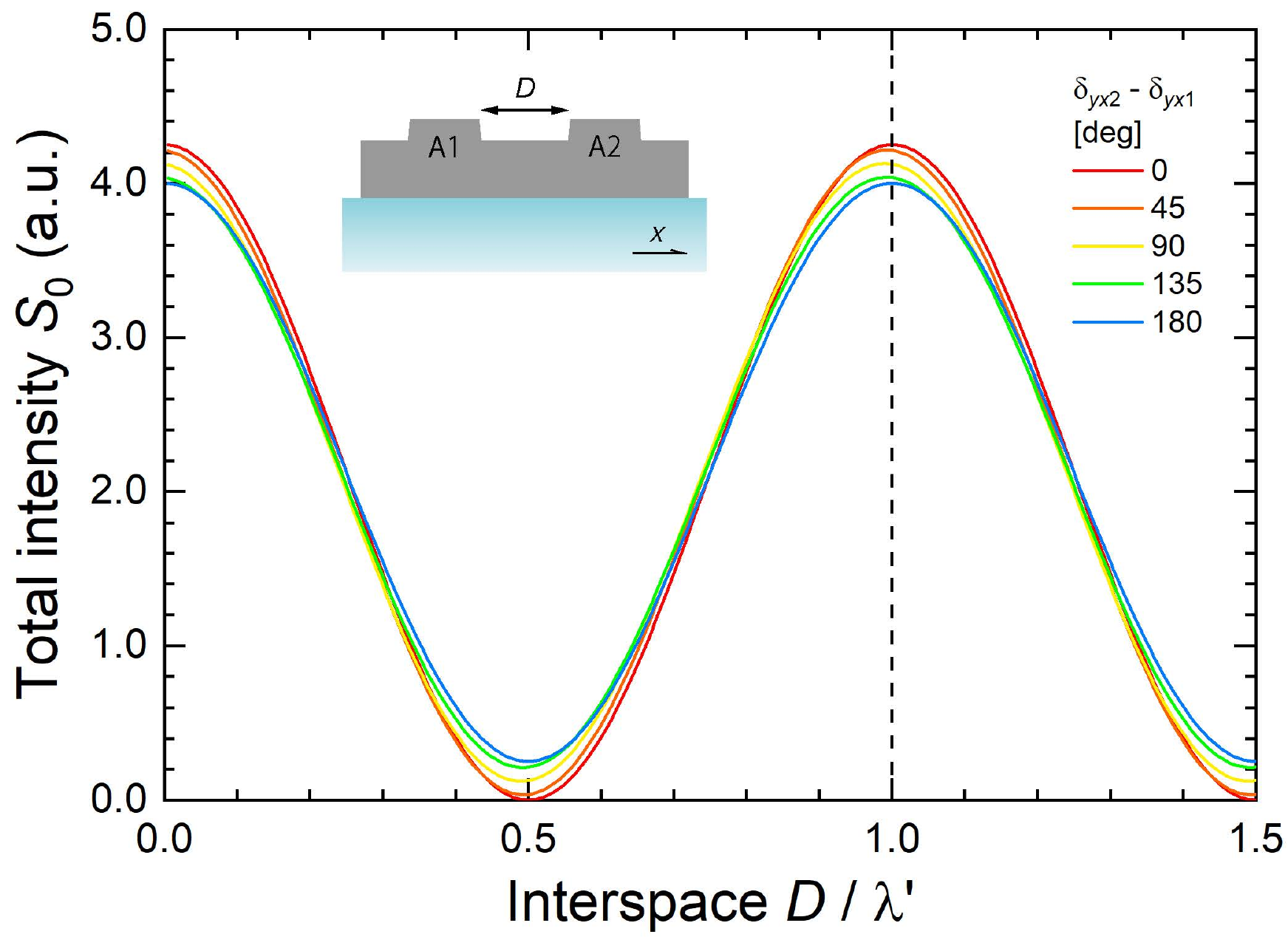} \\
	\caption{
Total intensity of the $E$-field emitted from two mesas as a function of the inter-mesa space $D$.  Inset: schematic side view of the Bi-2212 mesa array.
}
\end{figure}%